\documentstyle[twoside]{article}

\catcode`\@=11
\long\def\@makefntext#1{
\protect\noindent \hbox to 3.2pt {\hskip-.9pt  
$^{{\eightrm\@thefnmark}}$\hfil}#1\hfill}		

\def\thefootnote{\fnsymbol{footnote}}
\def\@makefnmark{\hbox to 0pt{$^{\@thefnmark}$\hss}}	
	
\def\ps@myheadings{\let\@mkboth\@gobbletwo
\def\@oddhead{\hbox{}
\rightmark\hfil\eightrm\thepage}   
\def\@oddfoot{}\def\@evenhead{\eightrm\thepage\hfil
\leftmark\hbox{}}\def\@evenfoot{}
\def\sectionmark##1{}\def\subsectionmark##1{}}

\oddsidemargin=\evensidemargin
\renewcommand{\thefootnote}{\fnsymbol{footnote}}

\newcounter{sectionc}\newcounter{subsectionc}\newcounter{subsubsectionc}
\renewcommand{\section}[1] {\vspace{12pt}\addtocounter{sectionc}{1} 
\setcounter{subsectionc}{0}\setcounter{subsubsectionc}{0}\noindent 
	{\tenbf\thesectionc. #1}\par\vspace{5pt}}
\renewcommand{\subsection}[1] {\vspace{12pt}\addtocounter{subsectionc}{1} 
	\setcounter{subsubsectionc}{0}\noindent 
	{\bf\thesectionc.\thesubsectionc. {\kern1pt \bfit #1}}\par\vspace{5pt}}
\renewcommand{\subsubsection}[1] {\vspace{12pt}\addtocounter{subsubsectionc}{1}
	\noindent{\tenrm\thesectionc.\thesubsectionc.\thesubsubsectionc.
	{\kern1pt \tenit #1}}\par\vspace{5pt}}
\newcommand{\nonumsection}[1] {\vspace{12pt}\noindent{\tenbf #1}
	\par\vspace{5pt}}

\newcounter{appendixc}
\newcounter{subappendixc}[appendixc]
\newcounter{subsubappendixc}[subappendixc]
\renewcommand{\thesubappendixc}{\Alph{appendixc}.\arabic{subappendixc}}
\renewcommand{\thesubsubappendixc}
	{\Alph{appendixc}.\arabic{subappendixc}.\arabic{subsubappendixc}}

\renewcommand{\appendix}[1] {\vspace{12pt}
        \refstepcounter{appendixc}
        \setcounter{figure}{0}
        \setcounter{table}{0}
        \setcounter{lemma}{0}
        \setcounter{theorem}{0}
        \setcounter{corollary}{0}
        \setcounter{definition}{0}
        \setcounter{equation}{0}
        \renewcommand{\thefigure}{\Alph{appendixc}.\arabic{figure}}
        \renewcommand{\thetable}{\Alph{appendixc}.\arabic{table}}
        \renewcommand{\theappendixc}{\Alph{appendixc}}
        \renewcommand{\thelemma}{\Alph{appendixc}.\arabic{lemma}}
        \renewcommand{\thetheorem}{\Alph{appendixc}.\arabic{theorem}}
        \renewcommand{\thedefinition}{\Alph{appendixc}.\arabic{definition}}
        \renewcommand{\thecorollary}{\Alph{appendixc}.\arabic{corollary}}
        \renewcommand{\theequation}{\Alph{appendixc}.\arabic{equation}}
        \noindent{\tenbf Appendix \theappendixc #1}\par\vspace{5pt}}
\newcommand{\subappendix}[1] {\vspace{12pt}
        \refstepcounter{subappendixc}
        \noindent{\bf Appendix \thesubappendixc. {\kern1pt \bfit #1}}
	\par\vspace{5pt}}
\newcommand{\subsubappendix}[1] {\vspace{12pt}
        \refstepcounter{subsubappendixc}
        \noindent{\rm Appendix \thesubsubappendixc. {\kern1pt \tenit #1}}
	\par\vspace{5pt}}

\newcommand{\textlineskip}{\baselineskip=13pt}
\newcommand{\smalllineskip}{\baselineskip=10pt}

\def\eightcirc{
\begin{picture}(0,0)
\put(4.4,1.8){\circle{6.5}}
\end{picture}}
\def\eightcopyright{\eightcirc\kern2.7pt\hbox{\eightrm c}} 

\newcommand{\copyrightheading}[1]
	{\vspace*{-2.5cm}\smalllineskip{\flushleft
	{\footnotesize International Journal of Modern Physics B, #1}\\
	{\footnotesize $\eightcopyright$\, World Scientific Publishing
	 Company}\\
	 }}

\newcommand{\publisher}[2]{{\begin{center}\footnotesize\smalllineskip 
	Received #1\\
	Revised #2
	\end{center}
	}}

\def\abstracts#1#2#3{{
	\centering{\begin{minipage}{4.5in}\baselineskip=10pt\footnotesize
	\parindent=0pt #1\par 
	\parindent=15pt #2\par
	\parindent=15pt #3
	\end{minipage}}\par}} 

\renewenvironment{thebibliography}[1]			
	{\frenchspacing
	 \ninerm\baselineskip=11pt
	 \begin{list}{\arabic{enumi}.}
	{\usecounter{enumi}\setlength{\parsep}{0pt}
	 \setlength{\leftmargin 12.7pt}{\rightmargin 0pt} 
	 \setlength{\itemsep}{0pt} \settowidth
	{\labelwidth}{#1.}\sloppy}}{\end{list}}

\newcounter{itemlistc}
\newcounter{romanlistc}
\newcounter{alphlistc}
\newcounter{arabiclistc}

\newcommand{\fcaption}[1]{
        \refstepcounter{figure}
        \setbox\@tempboxa = \hbox{\footnotesize Fig.~\thefigure. #1}
        \ifdim \wd\@tempboxa > 5in
           {\begin{center}
        \parbox{5in}{\footnotesize\smalllineskip Fig.~\thefigure. #1}
            \end{center}}
        \else
             {\begin{center}
             {\footnotesize Fig.~\thefigure. #1}
              \end{center}}
        \fi}

\newcommand{\tcaption}[1]{
        \refstepcounter{table}
        \setbox\@tempboxa = \hbox{\footnotesize Table~\thetable. #1}
        \ifdim \wd\@tempboxa > 5in
           {\begin{center}
        \parbox{5in}{\footnotesize\smalllineskip Table~\thetable. #1}
            \end{center}}
        \else
             {\begin{center}
             {\footnotesize Table~\thetable. #1}
              \end{center}}
        \fi}

\def\@citex[#1]#2{\if@filesw\immediate\write\@auxout
	{\string\citation{#2}}\fi
\def\@citea{}\@cite{\@for\@citeb:=#2\do
	{\@citea\def\@citea{,}\@ifundefined
	{b@\@citeb}{{\bf ?}\@warning
	{Citation `\@citeb' on page \thepage \space undefined}}
	{\csname b@\@citeb\endcsname}}}{#1}}

\newif\if@cghi
\def\cite{\@cghitrue\@ifnextchar [{\@tempswatrue
	\@citex}{\@tempswafalse\@citex[]}}
\def\citelow{\@cghifalse\@ifnextchar [{\@tempswatrue
	\@citex}{\@tempswafalse\@citex[]}}
\def\@cite#1#2{{$\null^{#1}$\if@tempswa\typeout
	{IJCGA warning: optional citation argument 
	ignored: `#2'} \fi}}

\def\pmb#1{\setbox0=\hbox{#1}
	\kern-.025em\copy0\kern-\wd0
	\kern.05em\copy0\kern-\wd0
	\kern-.025em\raise.0433em\box0}

\def\fnt#1#2{\footnotetext{\kern-.3em
	{$^{\mbox{\scriptsize #1}}$}{#2}}}

\def\fpage#1{\begingroup
\voffset=.3in
\thispagestyle{empty}\begin{table}[b]\centerline{\footnotesize #1}
	\end{table}\endgroup}

\def\runninghead#1#2{\pagestyle{myheadings}
\markboth{{\protect\footnotesize\it{\quad #1}}\hfill}
{\hfill{\protect\footnotesize\it{#2\quad}}}}
\font\tenrm=cmr10
\font\tenit=cmti10 
\font\tenbf=cmbx10
\font\bfit=cmbxti10 at 10pt
\font\ninerm=cmr9

\font\eightrm=cmr8

\def\qed{\hbox{${\vcenter{\vbox{			
   \hrule height 0.4pt\hbox{\vrule width 0.4pt height 6pt
   \kern5pt\vrule width 0.4pt}\hrule height 0.4pt}}}$}}

\renewcommand{\thefootnote}{\fnsymbol{footnote}}	

\def\bsc{{\sc a\kern-6.4pt\sc a\kern-6.4pt\sc a}}	
\def\bflatex{\bf L\kern-.30em\raise.3ex\hbox{\bsc}\kern-.14em 
T\kern-.1667em\lower.7ex\hbox{E}\kern-.125em X} 

\newcommand{\be}{\begin{equation}}
\newcommand{\ee}{\end{equation}}
\newcommand{\bea}{\begin{eqnarray}}
\newcommand{\eea}{\end{eqnarray}}
\newcommand{\non}{\nonumber}
\newcommand{\ra}{\rangle}

\newcommand{\lam}{\lambda} 
\newcommand{\Lam}{\Lambda} 
 
\newcommand{\al}{\alpha}

\begin{document}

\runninghead{The 8V CSOS model and the $sl_2$ loop algebra symmetry of the six-vertex model 
at roots of unity} 
{The 8VCSOS model and the $sl_2$ loop algebra symmetry of the six-vertex model at roots of unity}

\normalsize\textlineskip
\thispagestyle{empty}
\setcounter{page}{1}

\copyrightheading{}			

\vspace*{0.88truein}

\fpage{1}
\centerline{\bf The 8V CSOS model and the $sl_2$ loop algebra symmetry }
\vspace*{0.035truein}
\centerline{\bf of the six-vertex model at roots of unity}


\vspace*{0.37truein}
\centerline{\footnotesize TETSUO DEGUCHI \footnote{deguchi@phys.ocha.ac.jp}}
\vspace*{0.015truein}
\centerline{\footnotesize\it  Department of Physics, Ochanomizu University,
2-1-1 Ohtuska} 
\baselineskip=10pt
\centerline{\footnotesize\it Bunkyo-ku, Tokyo 112-8610,Japan} 

\vspace*{0.225truein}
\publisher{(received date)}{(revised date)}

\vspace*{0.21truein}
\abstracts{
We review  an algebraic method 
for constructing degenerate eigenvectors of  
the transfer matrix of the eight-vertex Cyclic Solid-on-Solid 
lattice model (8V CSOS model), 
where the degeneracy increases exponentially 
with respect to the system size. 
 We consider the elliptic quantum group $E_{\tau, \eta}(sl_2)$ 
at the discrete coupling constants: $2N \eta = m_1 + i m_2 \tau$, 
where $N, m_1$ and $m_2$ are integers.   
Then we show that degenerate eigenvectors 
of the transfer matrix of the six-vertex model 
at roots of unity in the sector $S^Z \equiv 0$ (mod $N$)  
are derived from those of the 8V CSOS model,  
through the trigonometric limit.  
They are associated with the complete $N$ strings. 
From the result we see that  
the dimension of a given degenerate eigenspace 
in the sector $S^Z \equiv 0$ (mod $N$) of the six-vertex model
at $N$th roots of unity 
is given by  $2^{2S_{max}^Z/N}$, where $S_{max}^Z$ is 
the maximal value of the total spin operator $S^Z$ 
in the degenerate eigenspace.  
}{}{}



\vspace*{1pt}\textlineskip	
\section{Introduction}	
\vspace*{-0.5pt}
\noindent
Recently, it has been explicitly discussed that 
the transfer matrix of the six-vertex model 
at roots of unity has the symmetry of the $sl_2$ loop algebra. 
\cite{DFM,FM1,FM2,FM3,TD}
Let us consider the  XXZ spin chain under the periodic boundary conditions  
\be 
H_{XXZ} = - J \sum_{j=1}^{L} \left(\sigma_j^X \sigma_{j+1}^X +
 \sigma_j^Y \sigma_{j+1}^Y + \Delta \sigma_j^Z \sigma_{j+1}^Z  \right) \, . 
\label{hxxz}
\ee
Here the parameter $\Delta$ is related to the $q$ variable of the quantum group $U_q(sl_2)$ as 
\be 
\Delta = {\frac 1 2} (q+ q^{-1}) 
\ee
When $q^{2N}=1$, it was shown \cite{DFM} that the XXZ Hamiltonian commutes with the 
generators of the $sl_2$ loop algebra, which is an infinite dimensional algebra. 
Furthermore, it was shown \cite{DFM} by the Jordan-Wigner method for $N=2$ and 
numerically for general $N$  that the dimensions of the degenerate eigenvectors 
are given by some powers of 2, which increase exponentially with respect to the system size $L$.

\par 
The exponential degeneracy of the $sl_2$ loop algebra 
should be important for the problem of the ``completeness 
of the Bethe ansatz eigenvectors''. In fact, 
the $sl_2$ loop algebra symmetry has not been considered  
in  the standard arguments of the string hypothesis.  
\cite{Takahashi,Kirillov} 
Thus, it seems that it is still open whether we can construct  
$2^L$ linearly independent eigenvectors of the XXZ spin chain  
at roots of unity for general $L$. The question should be 
related to so called singular Bethe ansatz solutions. \cite{Kim} 
In fact, it is numerically confirmed that 
the standard solutions of the Bethe ansatz equations determine 
only  eigenvectors which have 
the highest weights of the $sl_2$ loop algebra. \cite{FM1} 
Furthermore, some important properties of 
complete $N$ strings have been discussed  
in association with the $sl_2$ loop algebra. \cite{FM1,FM2,FM3} 
 
\par  
Interestingly, it was numerically suggested that 
the transfer matrix of the eight-vertex model 
at the discrete coupling parameters  
should have the degenerate eigenvectors  corresponding  to the degeneracy of 
 the $sl_2$ loop algebra. \cite{DFM} Furthermore, it has been recently shown  that 
 some degenerate eigenspace of the eight-vertex model 
 has dimension of $N 2^{L/N}$ if $L/N$ is an even integer. \cite{TD} 
 Let us consider the XYZ Hamiltonian under the periodic boundary conditions 
\cite{Baxter0,Baxter123} 
 \be 
H_{XYZ} = - \sum_{j=1}^{L} \left( J_X \sigma_j^X \sigma_{j+1}^X +
J_Y \sigma_j^Y \sigma_{j+1}^Y + J_Z \sigma_j^Z \sigma_{j+1}^Z  \right) 
\ee
where the coupling constants $J_X$, $J_Y$ and $J_Z$  are given by 
\be 
J_X  =  J(1 + k \, {\rm sn}^2( 2\eta) ) \, , \quad 
J_Y  =  J(1 - k \, {\rm sn}^2 ( 2\eta) ) \, , \quad 
J_Z  =  J \, {\rm cn}(2\eta) {\rm dn}(2\eta) 
\label{JXYZ}
\ee 
Here ${\rm sn}(z)$, ${\rm cn}(z)$ and  ${\rm dn}(z)$
 denote the Jacobian elliptic functions with elliptic modulus $k$.  
We have called $2\eta$ the coupling parameter of the model.  
The number $N$ has been related to $2\eta$ 
by $2N \eta = 2 m_1 K + i m_2 K^{'}$. 
The symbols $K$ and $K^{'}$ denote the complete elliptic integrals of the 
first and second kinds, respectively.

\par 
In this  paper, we discuss an algebraic construction of   
 degenerate eigenvectors of the eight-vertex cyclic 
Solid-on-Solid model \cite{Pearce,Kuniba,Akutsu}  (8V CSOS model), which 
is a variant of the eight-vertex Restricted Solid-on-Solid model (ABF model)
Then, we show that through some  
limits, they give the 
degenerate eigenvectors of the six-vertex model 
in the sector $S^Z \equiv 0$ (mod $N$) 
consisting of the complete $N$ strings.


\textheight=7.8truein
\setcounter{footnote}{0}
\renewcommand{\thefootnote}{\alph{footnote}}

\section{The $sl_2$ loop algebra symmetry of the XXZ spin chain}

Let us consider representations of the generators of $U_q(sl_2)$ on 
the $L$th tensor product of spin 1/2 representations.   
\begin{equation}
q^{S^Z}=q^{\sigma^Z/2}\otimes \cdots \otimes q^{\sigma^Z/2}
\end{equation}
\begin{equation}
S^{\pm}=\sum_{j=1}^LS_j^{\pm}=\sum_{j=1}^Lq^{\sigma^Z/2}\otimes
\cdots q^{\sigma^Z/2}\otimes\sigma_j^{\pm}\otimes q^{-\sigma^Z/2}\otimes
\cdots \otimes q^{-\sigma^Z/2}
\label{spm}
\end{equation}
Let us introduce some symbols: $[n]=(q^n-q^{-n})/(q-q^{-1})$ for $n>0$ 
and $[0]=1$;  ${[}n{]}! = \prod_{k=1}^n {[}k{]}$. Setting 
\begin{equation}
S^{\pm(N)}={\rm lim}_{q^{2N}\rightarrow 1}(S^{\pm})^N/[N]!
\label{limsn}
\end{equation}
the operators $S^{\pm(N)}$  are non-vanishing and we have 
\begin{eqnarray}
S^{\pm(N)}&=&  
\sum_{1 \le j_1 < \cdots < j_N \le L}
q^{{N \over 2 } \sigma^Z} \otimes \cdots \otimes q^{{N \over 2} \sigma^Z}
\otimes \sigma_{j_1}^{\pm} \otimes
q^{{(N-2) \over 2} \sigma^Z} \otimes  \cdots \otimes q^{{(N-2) \over 2}
\sigma^Z}
\nonumber \\
 & & \otimes \sigma_{j_2}^{\pm} \otimes q^{{(N-4) \over 2} \sigma^Z} \otimes
\cdots
\otimes \sigma^{\pm}_{j_N} \otimes q^{-{N \over 2} \sigma^Z} \otimes \cdots
\otimes q^{-{N \over 2} \sigma^Z}
\label{sn}
\end{eqnarray}

\par 
The study of the symmetries of the XXZ Hamiltonian under periodic boundary
conditions at roots of unity  was initiated in Ref. \cite{Pasquier}: 
 $S^{\pm(N)}$ commute with the Hamiltonian (\ref{hxxz}) 
 when $S^Z/N$ is an integer and $q^{2N}=1$ holds. 
However, there exists a much larger symmetry algebra than that of $S^{\pm(N)}$. 
\cite{DFM} We remark that the XXZ Hamiltonian is associated with the affine quantum group $U_q({\hat sl}_2)$.  
For instance, we may consider the following 
\begin{equation}
T^{\pm}=\sum_{j=1}^LT_j^{\pm}=\sum_{j=1}^Lq^{-\sigma^Z/2}\otimes
\cdots q^{-\sigma^Z/2}\otimes\sigma_j^{\pm}\otimes q^{\sigma^Z/2}\otimes
\cdots \otimes q^{\sigma^Z/2}
\label{tpm}
\end{equation}
which is also obtained from $S^{\pm}$ by the replacement $q\rightarrow
q^{-1}.$  When $q^{2N}=1$, we define $T^{\pm(N)}$ similarly as (\ref{limsn}). 

\par 
Let $T_{6V}(v)$ denotes the (inhomogeneous) transfer matrix of the six-vertex model.    
Then we can show the (anti) commutation relations when $S^Z \equiv 0 (\rm{mod}~N)$ \cite{DFM} 
\begin{equation}
S^{\pm (N)} T_{6V}(v)=q^N T_{6V}(v) S^{\pm (N)}, \qquad T^{\pm (N)} T_{6V}(v)=q^N T_{6V}(v) T^{\pm (N)} 
\end{equation}
and therefore in the sector $S^Z \equiv 0$ (mod $N$) we have 
\begin{eqnarray}
{[}S^{\pm(N)},H{]}={[}T^{\pm(N)},H{]}=0.
\label{sthcomm}
\end{eqnarray}

\par 
Let us discuss the symmetry algebra. With the following  identification  \cite{DFM} 
\begin{equation}
e_0=S^{+(N)}, \quad f_0=S^{-(N)}, \quad e_1=T^{-(N)}, \quad 
f_1=T^{+(N)},t_0=-t_1=-(-q)^NS^z/N \, , 
\end{equation}
we can show that they satisfy  the defining relations of the $sl_2$ loop algebra: 
\be
[S^{+(N)}, T^{+(N)}]  =  [S^{-(N)},T^{-(N)}] = 0 \, ,  \label{one}  
\ee
\be 
[S^{\pm (N)}, S^{Z}]  =  \pm N S^{ \pm (N)}, \qquad 
[T^{\pm (N)}, S^{Z}]= \pm N T^{\pm (N)} \, , \label{three} 
\ee
\begin{eqnarray}
S^{+(N)3}T^{-(N)}-3S^{+(N)2}T^{-(N)}S^{+(N)}+3S^{+(N)}T^{-(N)}S^{+(N)2}
-T^{-(N)}S^{+(N)3}&=&0 \non \\
S^{-(N)3}T^{+(N)}-3S^{-(N)2}T^{+(N)}S^{-(N)}+3S^{-(N)}T^{+(N)}S^{-(N)2}
-T^{+(N)}S^{-(N)3}&=&0 \non \\
T^{+(N)3}S^{-(N)}-3T^{+(N)2}S^{-(N)}T^{+(N)}+3T^{+(N)}S^{-(N)}T^{+(N)2}
-S^{-(N)}T^{+(N)3}&=&0 \non \\
T^{-(N)3}S^{+(N)}-3T^{-(N)2}S^{+(N)}T^{-(N)}+3T^{-(N)}S^{+(N)}T^{-(N)2}
-S^{+(N)}T^{-(N)3}&=&0 \, , \non \\
\end{eqnarray}
and in  the sector $S^z\equiv 0 ({\rm mod}~N)$ we have 
\begin{equation}
[S^{+(N)},S^{-(N)}]=[T^{+(N)},T^{-(N)}]=-(-q)^N{2\over N}S^z. \label{two}
\end{equation}
The loop algebras with higher ranks are also discussed for 
some vertex models. \cite{Korff}

\section{The algebraic Bethe ansatz of the elliptic quantum group $E_{\tau, \eta}(sl_2)$ }
\noindent
The elliptic algebra  $E_{\tau, \eta}(sl_2)$ 
 is an algebra generated by meromorphic functions of a variable 
 $h$ and the matrix elements of a matrix $L(z, \lam)$ 
 with non-commutative entries, \cite{FV1,FV2} 
 which satisfy the Yang-Baxter relation with a dynamical shift 
\bea 
& & R^{(12)}(z_{12}, \lam - 2\eta h^{(3)}) L^{(1)}(z_1,\lam) 
L^{(2)}(z_2,\lam - 2\eta h^{(1)}) \non \\
&=&  L^{(2)}(z_2, \lam) 
L^{(2)}(z_1, \lam- 2 \eta h^{(2)}) R^{(12)}(z_{12}, \lam)
\label{DYBR}
\eea
Here $h$ is a  generator of  the Cartan subalgebra ${\bf h}$ 
of $sl_2$. Drinfeld's quasi-Hopf algebra 
gives a natural framework for the dynamical Yang-Baxter relation,   
which  can be derived from the standard quantum group  
$U_q({\hat {sl}}_2)$ through the twist \cite{BBB,Jimbo}.

\par 
The $R$-matrix of (\ref{DYBR}) is essentially 
 that of the ABF model \cite{ABF} (the 8V RSOS model). 
Let $V$ be the two-dimensional complex vector space with 
the basis $e[1]$ and $e[-1]$. Here we denote $e[-1]$ also   
 as $e[2]$,  
and let $E_{ij}$ denote the matrix satisfying 
 $E_{ij} e[k] = \delta_{jk} e[i]$.  
Then, the $R$-matrix $R(z, \lam) \in End(V)$ is given by 
\bea 
& R(z,\lam; \eta, \tau) & =  E_{11} \otimes E_{11} + E_{22} \otimes E_{22} 
 + \alpha(z, \lam) E_{11} \otimes E_{22} \non \\
& + & \beta(z, \lam) E_{12} \otimes E_{21}  + \beta(z, -\lam) E_{21} \otimes E_{12}
    + \alpha(z, -\lam) E_{22} \otimes E_{11}
\label{R-matrix}
\eea
where $h=E_{11} - E_{22}$ and  
$\alpha(z, \lam)$ and $\beta(z, \lam)$ are defined by 
\be 
\alpha(z, \lam) = {\frac {\theta(z)\theta(\lam+2\eta)} {\theta(z-2\eta)\theta(\lam)}} \, , 
\qquad \beta(z, \lam) = - 
{\frac {\theta(z+\lam)\theta(2\eta)} {\theta(z-2\eta)\theta(\lam)}} \, . 
\ee
The theta function has been given by 
\be 
\theta(z; \tau) = 2 p^{1/4} \sin \pi z \prod_{n=1}^{\infty}
(1-p^{2n})(1-p^{2n}\exp(2\pi i z)) (1-p^{2n}\exp(-2\pi i z)) \, , 
\ee
where the nome $p$ is related to  the parameter $\tau$ by $p=\exp(\pi i \tau)$  
with  ${\rm Im} \quad \tau>0$ .

\par 
Let us now review the construction of the eigenvectors of the elliptic algebra $E_{\tau, \eta}(sl_2)$ 
at the discrete coupling parameter: $2N \eta = m_1 + m_2 \tau$ , where $N$, $m_1$ and $m_2$ 
are any given integers. \cite{TD} 
Here we note that  $2N \eta = m_1 + m_2 \tau$ corresponds to 
$2N \eta = 2 m_1 K + i m_2 K$ in (\ref{JXYZ}). 
Hereafter we assume $m_2=0$ for simplicity. 
Let $W=V(z_1) \otimes \cdots \otimes V(z_L)$ be the $L$th tensor product of 
the evaluation modules $V_{\Lam_j}(z_j)$'s with $\Lam_j=1$ for all $j$. \cite{FV1,FV2}  
The transfer matrix $T(z)$ of $E_{\tau, \eta}(sl_2)$ is given by 
the trace of the $L$-operator acting on the module $W$ 
\be 
 L(z, \lam)= R^{(01)}(z-z_1, \lam-2\eta \sum_{j=2}^{L} h^{(j)}) 
   R^{(02)}(z-z_2, \lam - 2\eta \sum_{j=3}^{L} h^{(j)}) 
   \cdots  R^{(0L)}(z-z_L, \lam)  
\label{L-SOS}
 \ee

\par Let us consider the $m$th product of the creation operators 
 $b(t_j)$'s  on the vacuum. \cite{8VABA,FV2}
Let us assume the number $m$ satisfies the following condition 
\be 
2m= L - rN \, , \quad \mbox{\rm for} \quad r \in {\bf Z}  
\ee
Hereafter we also assume that $r m_1$ is even. 
We introduce  a function $g_c(\lam)$ by 
 $g_c(\lam) = e^{c\lam} \, \prod_{j=1}^{m} \left( {\theta(\lam-2 \eta j)}/{\theta(2 \eta)} \right)$. 
Vector $v_c$ is defined by  $v_c= g_c(\lam) v_0$, 
where  $v_0$ is the highest weight vector of $W$: $h v_0 = L v_0$. 
Then,  making use of the fundamental commutation relations \cite{FV2} 
associated with $b(z_j)$'s, 
we can show that  
 $b(t_1) \cdots b(t_m) v_c$ 
 is an eigenvector of the transfer matrix 
$T(z)$ with the eigenvalue $C_0(z)$ 
\be 
C_0(w)  =  e^{-2 \eta c} \, \prod_{j=1}^{m} 
{\frac {\theta(w-t_j+2 \eta)}  {\theta(w-t_j)}}
+ e^{2 \eta c} \, \prod_{j=1}^{m} 
{\frac {\theta(w-t_j -2 \eta)}  {\theta(w-t_j)}}
\prod_{\al=1}^{L} 
{\frac {\theta(w- z_{\al})}  {\theta(w- z_{\al}- 2 \eta)}} \, , 
\label{eigenvalue} 
\ee
if rapidities $t_1, t_2, \ldots, t_m$ 
satisfy the Bethe ansatz equations  
\be 
\prod_{k=1}^{L} {\frac {\theta(t_j-p_k)} {\theta(t_j-q_k)}} 
=  e^{-4\eta c} \prod_{k=1; k \ne j}^{m} 
{\frac {\theta(t_j-t_k + 2\eta)} {\theta(t_j-t_k -2 \eta)}} 
\qquad {\rm for} \quad j=1, \ldots, m \, .
\label{BAE} 
\ee

\par 
The ``matrix elements'' of the vector $b(t_1) \cdots b(t_m) v_c$ is explicitly given  by \cite{FV2} 
\bea
& & b(t_1) \cdots b(t_m) v_c  =  (-1)^{m} e^{c(\lam + 2\eta m)} 
 \sum_{P \in {\cal S}_m} \sum_{1 \le j_1 < \cdots < j_m \le L} 
  \prod_{\al=1}^{m} \prod_{\beta =j_{\al}+1}^{L} 
  {\frac {\theta(t_{P\al} - z_{\beta})} {\theta(t_{P\al} - z_{\beta} - 2 \eta)} }  
   \non \\
&& \times  \, \prod_{1 \le \al < \beta \le m} f_{P\al P\beta} \, \times \,  
\prod_{\al=1}^{m} 
{\frac 
{\theta(\lam + t_{P \al} - z_{j_\al} - 2\eta(r N -j_{\al} + \al))}  
{\theta(t_{P\al} - z_{j_{\al}} - 2 \eta )} } 
\, \sigma^{-}_{j_1} \cdots \sigma^{-}_{j_m} \, |0 \ra 
\label{bbb}
\eea
Here $\sigma_{j}^{-}$ denotes the Pauli matrix $\sigma^{-}$ acting on 
the $j$th site, ${\cal S}$ the symmetric group, $| 0 \ra$ the vacuum vector 
and $f_{jk}= {\theta(t_j -t_k - 2 \eta)} /{\theta(t_j- t_k)}$.

\section{The eigenvectors of the 8V CSOS model}
\noindent
Let us replace $\lam$ with $\lam + \lam_0$ in the $L$-operator (\ref{L-SOS}) on $W$.  
Here $\lam_0$ is independent of $\lam$. 
Then, the  $R$-matrix $R(z, \lam+\lam_0)$    
is related to the Boltzmann weights  $w(a,b,c,d; z, \lam_0)$  of the 8V CSOS model 
 through the following relation 
\be
R(z, -2\eta d + \lam_0 ) e[c-d] \otimes e[b-c] = \sum_{a} w(a,b,c,d; z, \lam_0) e[b-a] \otimes e[a-d] 
\label{R-w}
\ee   
Here $a,b,c,d$ denote the spin variables of the IRF (the Interaction Round a Face) model 
which take integer values. \cite{Baxter123} 
The spin variables have  the constraint that 
the difference between the values of two nearest-neighboring spins should be  
given by  $\pm 1$.   
 Furthermore,  for the 8V CSOS model discussed in Refs.  \cite{Pearce,Kuniba,Akutsu},   
 the spin variables take the restricted values 
such as 0, 1, \ldots, $N-1$ where the values 0 and $N-1$ can be 
assigned for adjacent spins.

\par 
  Through the relation (\ref{R-w}), we can show that 
the transfer matrix $T(z)$ of $E_{\tau, \eta}(sl_2)$  
acting on the ``path basis'' corresponds 
to that of the 8V CSOS model \cite{Baxter123,FV2}.  
Here we note that a ``path'' is given by  
a sequence of spin  values  
satisfying the  constraints on adjacent spins.  
Explicitly we consider the following \cite{FV2}
\be 
|a_1, a_2, \cdots a_L \ra(\lam) = \delta(\lam+ 2 \eta a_1) \,  e[a_1 - a_2] \otimes 
e[a_2- a_3] \otimes \cdots 
\otimes e[a_L - a_1]
\ee
Here for the 8V CSOS model, we assume that $a_L - a_1 \equiv \pm 1$ (mod $N$). 
Expressing the eigenvector $b(t_1)\cdots b(t_m) v_c$ of $T(z)$  in terms of the path basis,  
we obtain  that of the transfer matrix of the 8V CSOS model.

\section{The degenerate eigenvectors 
of the transfer matrix of the 8V CSOS model}

\par 
Let us now assume that out of $m$ rapidities $t_1, \ldots, t_m$, 
the first $R$ rapidities $t_j$ for $j=1, \ldots, R$ are of standard ones 
 satisfying the Bethe ansatz equations (\ref{BAE}) with $m$ replaced by $R$,  
while  the remaining $N F$ rapidities are
  formal solutions given by  
\be 
t_{(\alpha, j)} = t_{(\alpha)} + \eta(2j-N-1) + 
\epsilon \, r_{j}^{(\alpha)}  \, , 
\qquad {\rm for} \quad j=1, \ldots, N\, . 
\ee 
We call the set of $N$ rapidities $t_{(\alpha, 1)}, \ldots, t_{(\alpha, N)}$, 
the complete $N$-string with  center $t_{(\alpha)}$. Here   
 the index $\al$ runs from 1 to $F$. 
Furthermore, we  assume that the index $(\al, j)$ corresponds to the number 
$R+ N (\al-1) +j$ for $1 \le \al \le F$ and $1 \le j \le N$. 
  We note that  the complete strings were suggested 
in Ref. \cite{Baxter123} in another context.  

\par 
Using the fundamental commutation relations,  we can show when $\epsilon \ne 0$
\bea 
& & T(z)b(t_1) \cdots b(t_{R+NF}) \, v_c  = 
C_0(z) b(t_1) \cdots b(t_{R+NF})\, v_c \non \\
&  & + \left( \sum_{j=1}^{R} +\sum_{j=R+1}^{R+NF} \right) \, C_{j} \, b(t_1) \cdots b(t_{j-1}) b(z) b(t_{j+1}) 
\cdots b(t_{R+NF}) \, v_c  \, . \label{Cbbb}
\eea
We  divide  eq. (\ref{Cbbb}) by $\epsilon$, 
  and send $\epsilon$ to zero.  
Then, we can show that each of the terms 
 of eq. (\ref{Cbbb}) indeed converges, 
by making use of the following formula 
\be 
 \prod_{1 \le \al< \beta \le m} f_{P \al P \beta } 
= \prod_{1 \le \al< \beta \le m } f_{\al \beta} \, 
 \, \times \, 
\prod_{1 \le j < k \le m} 
 \left(   {\frac {\theta(t_j-t_k + 2\eta)}{\theta(t_j-t_k - 2\eta)}}
  \right)^{H(P^{-1}j - P^{-1}k)}  , 
\label{formula} 
\ee
for  $P \in {\cal S}_m$.  
Here  $H(x)$ denotes the Heaviside step function: $H(x)=1$ for $x> 0$,  
$H(x) = 0$ otherwise. The symbol  $P \in {\cal S}_m$ 
denotes an element $P$ of the  symmetric group of $m$ elements,  
where $j$ is sent to  $Pj \in \{ 1, 2, \ldots, m \}$ for $j=1, \ldots m$.  
The formula (\ref{formula}) has been  proven in Ref. \cite{XXX}. 

\par 
Let us consider the following  function of variable $z$ \cite{TD} 
\be 
G(z)  = 
 \sum_{a=1}^{N}
 e^{-4 \eta c a} \prod_{j=a+1}^{N} 
\prod_{k=1}^{R} {\frac {\theta(z-t_k + \eta(2j -N+1) )} {\theta(z-t_k + \eta(2j -N-3) )} } 
\prod_{\beta=1}^{L} {\frac {\theta(z- z_{\beta} + \eta(2j -N-3) )} 
 {\theta(z- z_{\beta} + \eta(2j -N-1))} }  
\ee
Hereafter we assume $\exp(4N \eta c )=1$. 
Then, the centers $t_{(\al)}$ 's are determined  by  
 \be 
 G(z=t_{(\al)})= 0 \,  , \quad {\rm for } \quad \al=1, \ldots, F \, . 
\label{GGG} 
 \ee 
We can show that the zeros of (\ref{GGG}) also form complete $N$ strings, 
and also that the number of zeros of (\ref{GGG}) is given by $L- 2R$, 
by using the Bethe ansatz equations (\ref{BAE}). \cite{TD} Thus, 
the number of independent solutions to (\ref{GGG}) is given by 
$(L-2R)/N$, which leads to the dimension $2^{(L-2R)/N}$ through the binomial expansion.     
Thus, for the transfer matrix of the 8V CSOS model, 
 any standard  Bethe ansatz eigenvector with $R$ rapidities 
has the degeneracy of $2^{(L-2R)/N}$.

\par 
Let us now consider the connection of the CSOS model to the six-vertex model.   
Taking the trigonometric limit: $\tau \rightarrow i \infty$  and 
sending $\lam_0$ to infinity with  some gauge transformations, 
the $L$-operator of the 8V CSOS model becomes that of the six-vertex model. 
We may assume that the 
trigonometric limits of the $R$ rapidities of the Bethe ansatz equations (\ref{BAE}) 
with $\exp(4 \eta c)=1$  satisfy the trigonometric Bethe ansatz equations of the
six-vertex model. Then,  the degenerate eigenvectors with $F$ complete $N$ strings 
for the 8V CSOS model 
become those of the six-vertex model with $F$ complete $N$ strings. 
Thus, we have shown that the corresponding degenerate eigenspace is 
spanned by the eigenvectors having complete $N$ strings, 
and also that  the dimension is given by  $2^{(L-2R)/N}=2^{2 S_{max}^Z/N}$ since 
the highest weight $S^Z_{max}$ is given by $L/2 -R$.   
The result should be consistent   
with the previous studies. \cite{DFM,FM1,FM2,FM3}

\nonumsection{Acknowledgements}
\noindent

The author would like to thank  Prof. Y. Akutsu, Prof. K. Fabricius and Prof. B.M. McCoy 
for helpful discussions and  valuable comments.  
He is  thankful to Prof. M.L. Ge for his kind invitation to the workshop ``Nankai Symposium'', 
 October 8-11, 2001,  Tianjin, China.    
This work is partially supported by the Grant-in-Aid for Encouragement 
of Young Scientists (No. 12740231).

\nonumsection{References}



\noindent

\begin{thebibliography}{000}


\bibitem{DFM} T. Deguchi, K. Fabricius and B.M. McCoy, J. Stat. 
Phys. {\bf 102}, 701 (2001).   
\bibitem{FM1} K. Fabricius and B.M. McCoy, 
J. Stat. Phys. {\bf 103}, 647 (2001). 
\bibitem{FM2} K. Fabricius and B.M. McCoy, J. Stat. Phys. {\bf 104}, 575 (2001). 
\bibitem{FM3} K. Fabricius and B.M. McCoy,  cond-mat/0108057. 
\bibitem{TD} T. Deguchi, cond-mat/0109078. 
%

\bibitem{Takahashi} M. Takahashi and M. Suzuki, Prog. of Theor. Phys. 
{\bf 46}, 2187 (1972). 
\bibitem{Kirillov} A.N. Kirillov 
and N.A. Liskova, J. Phys. A {\bf 30}, 1209 (1997). 

\bibitem{Kim} J.D. Noh, D.-S. Lee and D. Kim, Physica A {\bf 287}, 167 (2000). 


\bibitem{Baxter0} Baxter, Ann. Phys. {\bf 70}, 193 (1972). 
\bibitem{Baxter123} R. Baxter, Ann. Phys. {\bf 76}, 1 (1973);  {\bf 76}, 25 (1973); 
 {\bf 76}, 48 (1973).


\bibitem{Pearce} P.A.  Pearce and K.A. Seaton, Phys. Rev. Lett. {\bf 60}, 1347 (1988). 
\bibitem{Kuniba} A. Kuniba and T. Yajima, J. Stat. Phys. {\bf 52}, 829 (1987). 
\bibitem{Akutsu} Y. Akutsu, T. Deguchi and M. Wadati, J. Phys. Soc. Jpn. {\bf 57}, 1173 (1988). 

\bibitem{ABF} G.E. Andrews, R.J. Baxter and P.J. Forrester, 
J. Stat. Phys. {\bf 35}, 193 (1984). 


\bibitem{Pasquier} V. Pasquier and H. Saleur, 
Nucl. Phys. B330, 523 (1990).

\bibitem{Korff} C. Korff and B.M. McCoy, hep-th/0104120. 





\bibitem{FV1} G. Felder and A. Varchenko, Commun. Math. Phys. {\bf 181}(1996) 741 . 
\bibitem{FV2} G. Felder and A. Varchenko, Nucl. Phys. B {\bf 480} (1996) 485. 


\bibitem{BBB} O. Babelon, D. Bernard, E. Billey, 
Phys. Lett. B {\bf 375}, 89 (1996).  
\bibitem{Jimbo} M. Jimbo, H. Konno, S. Odake, J. Shiraishi, 
q-alg/9712029.   


\bibitem{8VABA} L. Takhtajan and L. Faddeev, 
 Russ. Math. Survey {\bf 34}(5), 11 (1979). 




\bibitem{XXX} T. Deguchi, cond-mat/0107260, to appear in J. Phys. A.  

%
%

\end{thebibliography}
\end{document}